%% file: conference_101719.tex
\documentclass[conference]{IEEEtran}
\input{preamble/header}
\input{acronyms/acronyms}
\usepackage{eso-pic}
\usepackage{url}
\AddToShipoutPictureBG*{
  \AtPageUpperLeft{%
    \put(0,-40){\raisebox{15pt}{\makebox[\paperwidth]{\begin{minipage}{21cm}\centering
      \textcolor{gray}{This article has been accepted for publication in the proceedings of the \\2021 IEEE International Conference on Wireless and Mobile Computing, Networking And Communications (WiMob).\\10.1109/WiMob52687.2021.9606272} 
    \end{minipage}}}}%
  }
  \AtPageLowerLeft{%
    \raisebox{25pt}{\makebox[\paperwidth]{\begin{minipage}{21cm}\centering
      \textcolor{gray}{ \copyright 2021 IEEE.  Personal use of this material is permitted.  Permission from IEEE must be obtained for all other uses, in any current or future media, including reprinting/republishing this material for advertising or promotional purposes, creating new collective works, for resale or redistribution to servers or lists, or reuse of any copyrighted component of this work in other works.
      }
    \end{minipage}}}%
  }
}

\begin{document}
\input{content/00_01_title}
\input{content/00_02_abstract}

\input{content/01_intro}
\input{content/02_background}
\input{content/03_system_architecture}
\input{content/04_algorithm}
\input{content/05_evaluation}
\input{content/06_conclusion}

\bibliographystyle{IEEEtran}
\bibliography{bib/IEEEabrv, bib/references}
\end{document}

%% file: preamble/header.tex
\usepackage{cite}
\usepackage{amsmath,amssymb,amsfonts}
\usepackage{algorithmic}
\usepackage{graphicx}
\usepackage{textcomp}
\usepackage{xcolor}
\def\BibTeX{{\rm B\kern-.05em{\sc i\kern-.025em b}\kern-.08em
    T\kern-.1667em\lower.7ex\hbox{E}\kern-.125emX}}

\usepackage{preamble/orcidlink}
\usepackage[nolist,nohyperlinks]{acronym}
\usepackage[binary-units, detect-weight=true, detect-family=true]{siunitx}
\DeclareSIUnit{\belmilliwatt}{Bm}
\DeclareSIUnit{\bel}{B}

%% file: acronyms/acronyms.tex
\begin{acronym}
    \acro{IPS}{Indoor Positioning System}
    \acro{IoT}{internet of things}
    \acro{BLE}{Bluetooth Low Energy}
    \acro{RSSI}{received signal strength indicator}
    \acro{kNN}{k-Nearest Neighbors}
    \acro{wkNN}{weighted k-Nearest Neighbors}
    \acro{WLAN}{Wireless Local Area Network}
    \acro{UWB}{Ultra-Wideband}
    \acro{GPS}{Global Positioning System}
    \acro{RFID}{radio-frequency identification}
    \acro{CNN}{convolutional neural network}
    \acro{CPU}{central processing unit}
    \acro{GPU}{graphics processing unit}
    \acro{MCU}{microcontroller unit}
    \acro{SoC}{system on a chip}
    \acro{SDK}{software development kit}
    \acro{GAP}{Generic Access Profile}
    \acro{GATT}{Generic Attribute Profile}
    \acro{USB}{Universal Serial Bus}
    \acro{UART}{Universal Asynchronous Receiver Transmitter}
    \acro{LoRa}{Long Range}
\end{acronym}

%% file: content/00_01_title.tex
\title{Design and Implementation of an RSSI-Based Bluetooth Low Energy Indoor Localization System}

\author{\IEEEauthorblockN{Silvano Cortesi}
\IEEEauthorblockA{\textit{Dept. of ITET} \\
\textit{ETH Zürich}\\
Zürich, Switzerland \\
\orcidlink{0000-0002-2642-0797}}
\and
\IEEEauthorblockN{Marc Dreher}
\IEEEauthorblockA{\textit{Dept. of MAVT} \\
\textit{ETH Zürich}\\
Zürich, Switzerland \\
\orcidlink{0000-0001-6617-9996}}
\and
\IEEEauthorblockN{Michele Magno}
\IEEEauthorblockA{\textit{Dept. of ITET} \\
\textit{ETH Zürich}\\
Zürich, Switzerland \\
\orcidlink{0000-0003-0368-8923}}
}

\maketitle

%% file: content/00_02_abstract.tex
\begin{abstract}
\Ac{IPS} is a crucial technology that enables medical staff and hospital managements to accurately locate and track persons or assets inside the medical buildings. Among other technologies, \ac{BLE} can be exploited for achieving an energy-efficient and low-cost solution. This work presents the design and implementation of an \ac{RSSI}-based indoor localization system. The paper shows the implementation of a low complex weighted k-Nearest Neighbors algorithm that processes raw \ac{RSSI} data from connection-less iBeacon’s. The designed hardware and firmware are implemented around the low-power and low-cost nRF52832 from Nordic Semiconductor. Experimental evaluation with the real-time data processing has been evaluated and presented in a \SI{7.2}{\meter} by \SI{7.2}{\meter} room with furniture and 5 beacon nodes. The experimental results show an average error of only \SI{0.72}{\meter} in realistic conditions. Finally, the overall power consumption of the fixed beacon with a periodic advertisement of \SI{100}{\milli\second} is only \SI{50}{\micro\ampere} at \SI{3}{\volt}, which leads to a long-lasting solution of over one year with a \SI{500}{\milli\ampere{}\hour} coin battery.
\end{abstract}

\acresetall

%% file: content/01_intro.tex
\section{Introduction}
    The complexity of medical buildings such as hospitals, clinics and nursing facilities are increasing every day due to the increasing number of patients/hosts~\cite{1}. Moreover, the complexity increases due to new equipment and technologies offered to the staff, ranging from medical equipment to electronic medical data storage. In recent years, we are assisting with the introduction of information technology location aware devices and systems for healthcare environments that are exploiting different \ac{IoT} technologies~\cite{1},~\cite{2}. The main goal of those emerging \ac{IPS} is enabling medical staff and hospital managements to accurately locate and track persons or assets inside the medical buildings~\cite{1}. Among other scenarios where \ac{IPS} can be exploited, the promising hospital scenarios could become a reality for \ac{IPS} in the near future: the vision is to have a next-generation nurse smart-calling system that is able to locate the nearest nurse in the medical building. This can make medical work more efficient~\cite{3}, e.g. to locate and track medical equipment, which simultaneously helps identifying paths of infection~\cite{4}. Moreover, finding equipment and other medical devices inside a hospital or a building is often a complicated and time-consuming task. In large hospitals time is crucial and having a fast and reliable way to identify the location of equipment could save lives~\cite{6}. Finally the \ac{IPS} leads to more freedom for patients and helps medical staff to track patients inside the hospital or large indoor areas. This can especially be helpful when patients with a critical disease such as dementia get lost in the medical building. These are only a few examples of how an \ac{IPS} can improve the internal functionalities inside a hospital.
    \par
    Advances in electronics, computer sciences, miniaturization, and wireless communication enable energy-efficient communication among battery-operated smart devices inside the \ac{IoT} paradigm~\cite{7}~\cite{8}. The presence of a smart device allowing the discovery, communication, and data analysis, is the key enabling factor for the real development and deployment of \ac{IoT} applications such as \ac{IPS}~\cite{1}. Wireless communication technologies are enabling \ac{IPS} with a precision of a few meters~\cite{9}. The development of a functional localization system accurate to one or two meters could be applied to many application scenarios, especially within the medical section. In addition to the highest possible precision, this should also be cost-effective and easy to set up as possible which guarantees a large application area. \ac{GPS} is a well-known and established technology to precisely identify the location – however, it is well-known that it cannot be used indoors~\cite{9}~\cite{10}. Different systems and technologies have been investigated and proposed in the recent literature, and few of them are available on the market to address the need for indoor localization, including systems that are using cameras or other combinations of sensors~\cite{11}.  However, emerging research is investigating new ways using wireless communication to enable accurate indoor localization~\cite{9}~\cite{12}.
    \par
\begin{figure}[htbp!]
        \centerline{\includegraphics[width=\columnwidth]{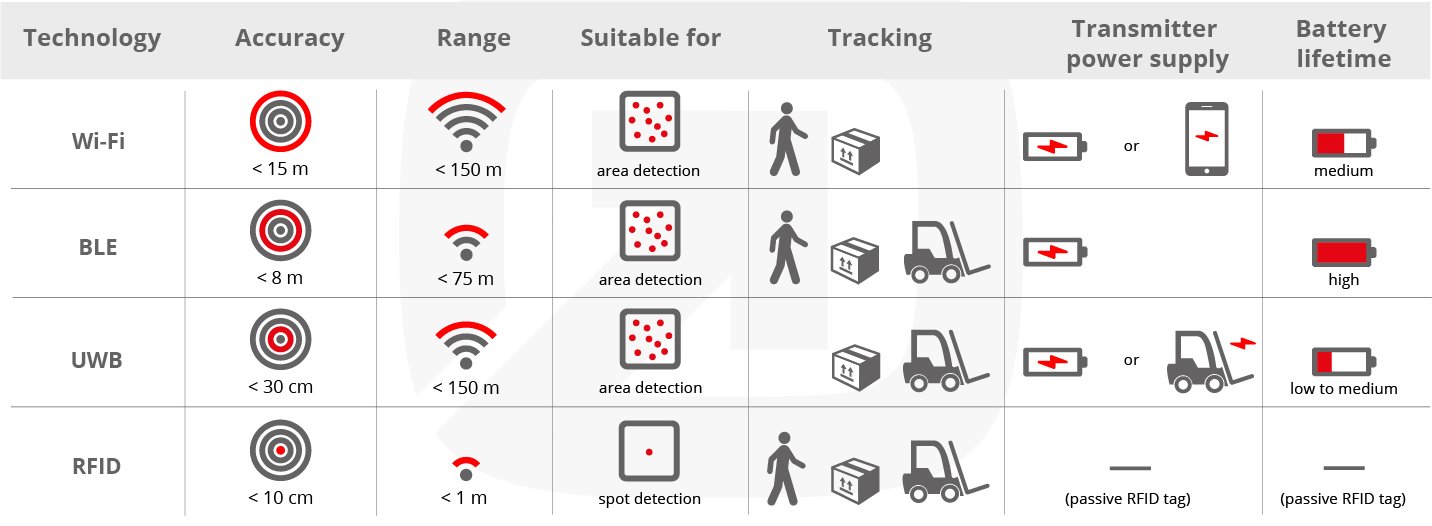}}
        \caption{The most common wireless technologies exploited for localization in buildings, \ac{LoRa} and \ac{GPS} are excluded since unsuitable accuracy in indoor application scenarios.}
        \label{technology_comparison}
    \end{figure}

    The most promising wireless technologies for meters-precision indoor localization are the \ac{WLAN}~\cite{13}, \ac{UWB}~\cite{14}-~\cite{16}, \ac{RFID}~\cite{17}, \ac{LoRa}~\cite{18} including the combination with \ac{GPS}~\cite{19}, and \ac{BLE}. Many researchers have been investigating the possibility of using \ac{WLAN} for the fact that it is pervasive in all buildings, and it has a long-range – however, the high energy consumption for a battery-powered system and the not excellent accuracy in the detection are the main weaknesses. Due to the ultra-wideband and the nature of the technology~\cite{14}, the solutions based on \ac{UWB} allow very high accuracy of less than \SI{30}{\centi\meter} in a radius of hundreds of meters - however current \ac{UWB} modules have still a too high power consumption in the range of hundreds of \si{\milli\watt}~\cite{16} that reduces its use in a real-application scenario where the \ac{IoT} device needs to last months with a coin battery. \ac{RFID} allows to achieve very low energy consumption as many tags can even work without any supply. Moreover, the passive tags achieve the most accurate \si{\centi\meter} accuracy. The operative range is only to the small range of less than 1m while the active tags for extending the ranges are affected by larger energy consumption and bigger antennas. Long Range sub-giga wireless communication such as \ac{LoRa}~\cite{20} have been investigated to provide localization information, however, the range achieved is in the order of several hundred meters, which makes them not suitable for high-precision indoor localization. Some researchers combined \ac{LoRa} with \ac{GPS} to achieve below-meter accuracy with real-time Kinematics, however, this solution works only outdoor~\cite{19}. \Acf{BLE} 5.x offers an optimal trade-off of accuracy, range, energy consumption, and availability in all modern phones that has been exploited by many recent works, including the novel features of estimation of angle of arrival enabled from the 5.2 version of \ac{BLE}~\cite{21}. Fig.~\ref{technology_comparison} gives an overview of the most exploited wireless technologies for indoor localization with some key parameters such as range, accuracy and energy efficiency when supplied by batteries.
    \par
    Although technology is advancing wireless communications, indoor localization systems are still prone to errors - attenuated by software algorithms. Many previous works focusing on the use of \acp{CNN} have been proposed as a competitive solution, especially for \ac{WLAN}~\cite{23} and \ac{BLE}~\cite{24}. \Acp{CNN} are improving the precision performance on one hand and on the other hand, \ac{CNN} models are designed to be executed on a \acs{CPU} or \acs{GPU}, requiring data to be transmitted from the mobile sensor node to an external compute engine through wired or wireless communication. Recently, a new generation of mobile smart \ac{IoT} are attracting academic and industrial researchers, using \acp{MCU}, supposed to bring computing capabilities towards the “edge” to perform real-time computation~\cite{25}. Edge computing offers the following advantages: 1) lower energy consumption for the data transmission between IoT devices and remote processing; 2) longer battery lifetime; 3) significantly shorter latency compared to remote computation; 4) user comfort; 5) security and privacy improvements, as the data are processed locally~\cite{25}.
    \par
    This paper focuses on the design and implementation of a hardware-software low-cost and battery-operated system for indoor localization. In particular, the system is based on \ac{BLE} 5.0 using the low-power general-purpose multi-protocol \ac{SoC} nRF52832 from Nordic Semiconductor.
    We hereby present the following contributions:
    \begin{itemize}
        \item On the software side, the paper proposes a system which is able to work with packets received via \ac{BLE} and an algorithm based on (w)\ac{kNN}.
        \item The design of the anchors and the mobile nodes are presented and evaluated with experimental results showing an overall accuracy of the implemented algorithm below \SI{1}{\meter} on average for a \(\SI{7.2}{\meter}\times\SI{7.2}{\meter}\) room.
    \end{itemize}

%% file: content/02_background.tex
\section{Background on \acs{RSSI}-Based \acs{BLE} Localization}
\begin{figure}[htbp!]
        \centerline{\includegraphics[width=0.8\columnwidth]{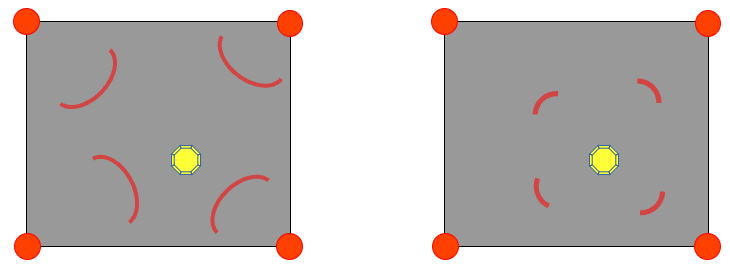}}
        \caption{Two possible constellations for localization based on \acs{RSSI} using \acs{BLE}.}
        \label{localization_variants}
    \end{figure}

    Localization with \ac{BLE} based on \ac{RSSI} can be implemented in two ways. One corresponds to the variant in which a mobile device receives packets from the fixed anchors and records the signal strength. The other method would be to have multiple fixed receivers and a mobile device as a transmitter of the packets. Fig.~\ref{localization_variants} illustrates the two options where the yellow circle in the middle identifies the mobile tag. The 4 circles in the corners are the \ac{BLE} anchors that are fixed.
    \par
    The former configuration achieves a higher energy efficiency of the mobile device, as it can wake up autonomously and decide when to receive messages and must not be connected to the anchors. In addition, all calculations can be carried out on the receiver – thus the beacon themselves can be passive without further communication. The fixed beacons, also called anchors, broadcast their unique identifier number. Moreover the mobile device receives these packets with their \ac{RSSI} and processes the data. To recognize the position correctly, training data must be collected at certain locations before usage. This training data is then like a footprint, which can be compared with the measured data and thus indicates the respective position.

    \subsection{\acs{BLE} and iBeacon Protocol}
        As protocol for the beacon, the decision has been made on the proprietary standard for indoor localization using \acs{BLE} iBeacon~\cite{26} from Apple because this protocol is commonly used in many applications and commercial systems, and it is fully supported by the Nordic \ac{SDK}~\cite{27} and other commercial \acp{SDK} of \ac{BLE} modules. To understand how the beacons work, the principle of \ac{BLE} should be clear first. In \ac{BLE}, there are two different operating modes~\cite{28}: \Ac{GAP} and \ac{GATT}. The \ac{GAP} is responsible for ensuring that the device is visible from the outside.
        \par
        \textit{\acs{GAP}:} In \ac{GAP}, the involved devices are not connected to each other. There are different roles and the most frequently used are central and peripheral. Simply put, peripherals behave like servers and centrals like clients. This means peripherals are devices that provide data. These can be beacons, heart-rate monitors, or other devices. Centrals are usually smartphones, tablets, or computers. In GAP, data can be transferred in two ways. First, there is the so-called Advertising Data payload, which is sent by the peripheral in constant time intervals to all clients (broadcasting). Then there is the Scan Response payload, which the central can request from the peripheral if it wants to learn more about the peripheral. Both packets are \SI{31}{\byte} in size, but only the Advertising Data payload must be implemented. The time between two advertisements can be freely chosen (of course, in compliance with the \ac{MCU} limits). Mostly the \ac{GAP} mode is only used so that other centrals can recognize the peripheral, and they can connect (which allows a faster and bidirectional transmission and enables specific services). Another use is to broadcast the same message to different devices. Once a connection has been established between a peripheral and a central, they switch to the \ac{GATT} mode.
        \par
        \textit{\acs{GATT}:} The \ac{GATT} defines the way of how two connected devices exchange their data. Typically, the systems look at the peripheral as a \ac{GATT} server, while the central one provides the \ac{GATT} client, making requests to the server. In the \ac{GATT}, only the central can initialize a transaction.  The \ac{GATT} transactions are based on Profiles, Services, and Characteristics.
        \par
\begin{figure}[htbp!]
            \centerline{\includegraphics[width=0.9\columnwidth]{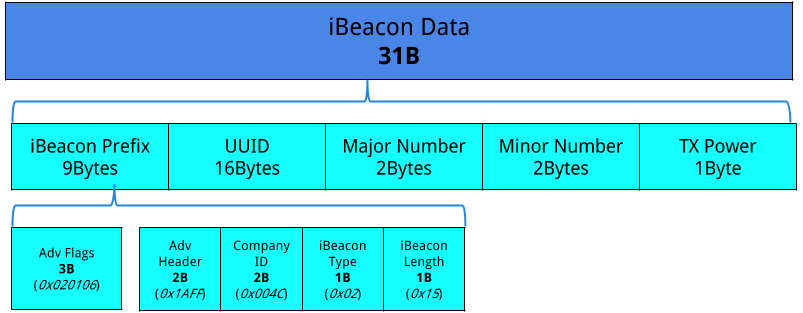}}
            \caption{Structure of iBeacon~\cite{31}}
            \label{ibeacon}
        \end{figure}

        \textit{iBeacon Protocol:} Since beacons do not want to be connected to a specific device but send their packets to all possible devices, they are in \ac{GAP} mode. This means that their packets can have a maximum size of \SI{31}{\byte} and they advertise their packets at constant intervals. The iBeacon's structure is presented in Fig.~\ref{ibeacon} and reported in~\cite{31}.
        \Ac{BLE} hardware module can identify the \ac{RSSI} of each received iBeacon, and this is the key information used in \ac{RSSI}-based localization systems, as the \ac{RSSI} can estimate the transmitter’s distance. 

    \subsection{Fading of \acl{RSSI}}
        To enable \ac{RSSI}-based location, the decrease in signal strength has to be clarified with the increase in distance in a medium such as air. This information can then be used in a trilateration algorithm (or to increase robustness of the \ac{kNN} further) to estimate a distance based on signal strength.  A well-known mathematical model is the logarithmic distance loss model. In this model, the decrease in signal strength is assumed to be logarithmic decreasing with the increasing distance~\cite{32}. Equation (\ref{eq:rssi}) illustrates the model used for the \ac{RSSI} value to approximate the distance.
        \begin{equation}\label{eq:rssi}
            \mathit{RSSI}(d)=\mathit{RSSI}(d_0)-10n\cdot\log_{10}\left(\frac{d}{d_0}\right)
        \end{equation}
        Where \(d\) is the estimated distance, \(d_0\) describes a self-defined reference distance, \(\mathit{RSSI}()\) the function that assigns the expected signal strength to each distance and \(n\) the path loss parameter given by the properties of the environment.
        \par
\begin{figure}[htbp!]
            \centerline{\includegraphics[width=0.9\columnwidth]{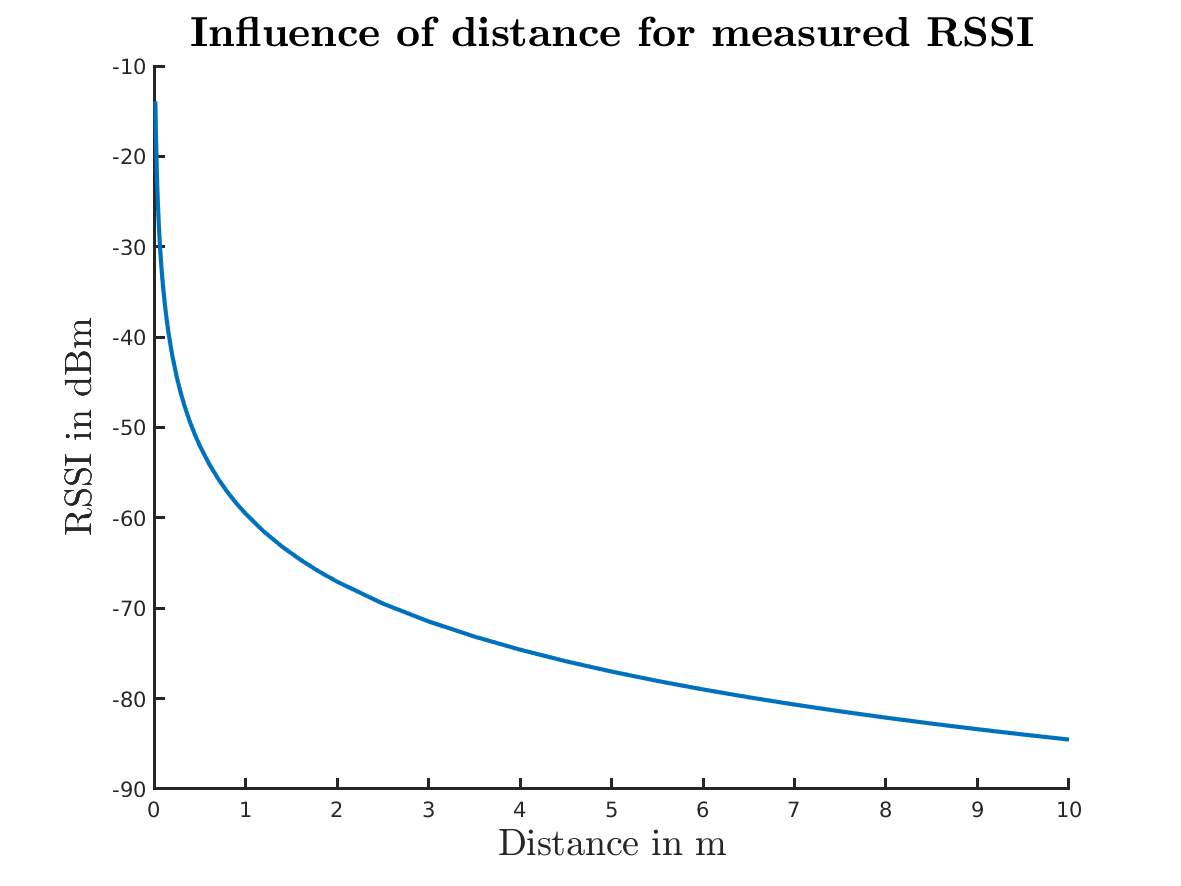}}
            \caption{Theoretical path-loss decay calculated for the Nordic nRF52832 used in this work.}
            \label{rssi_vs_d}
        \end{figure}

        Fig.~\ref{rssi_vs_d} shows the decay evaluated with the equation (\ref{eq:rssi}) based on approximations done in a room with the received \ac{RSSI} values of the nRF52832 for both beacon and receiver and transmission power of \SI{0}{\deci\belmilliwatt}.

%% file: content/03_system_architecture.tex
\section{System Architecture}
    This paper presents both the beacons and the receiver that were developed based on the \acl{BLE} 5.0 \ac{SoC} nRF52832 from Nordic Semiconductor. This was the most advanced \ac{BLE} chip at the beginning of the design, as it had the lowest power consumption among all available chips. This module contained an ARM Cortex-M4F microcontroller for the user firmware. 

    \subsection{Beacon nodes}
\begin{figure}[htbp!]
            \centerline{\includegraphics[width=0.8\columnwidth]{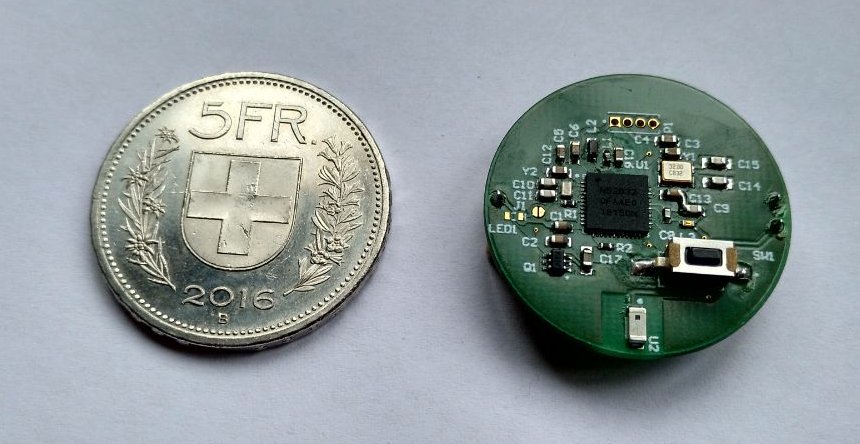}}
            \centerline{\includegraphics[width=0.8\columnwidth]{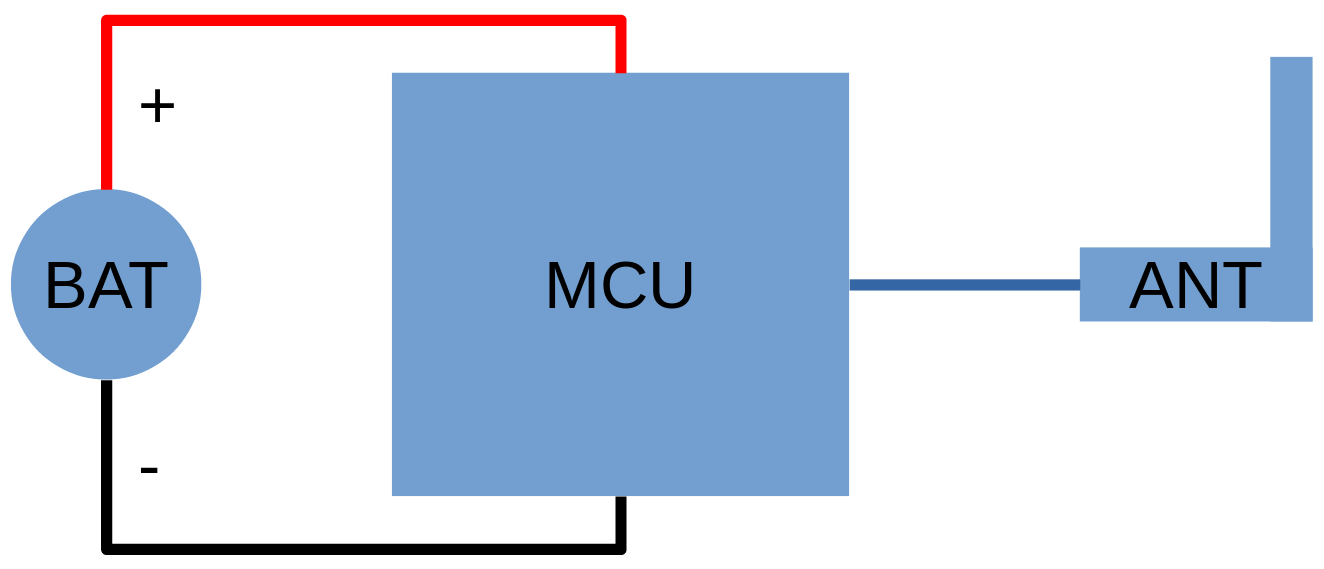}}
            \caption{Prototype of the beacon and block diagram.}
            \label{beacon}
        \end{figure}

        Fig.~\ref{beacon} shows the prototype and block diagram of the beacon developed and implemented in this work. The beacon is designed to operate from a \SI{3}{\volt} button battery, which is used directly to power the nRF52832, represented by the \ac{MCU} block in the figure. The beacon is designed for low power consumption and a small footprint (limited by the battery holder, the diameter is now \SI{3}{\centi\meter}). Power consumption averages \SI{50}{\micro\ampere}, resulting in over a year of runtime on a \SI{500}{\milli\ampere{}\hour} battery, even taking into account battery self-discharge. Due to the low power consumption, the beacon node could easily be powered only by photovoltaic energy harvesting allowing an unlimited runtime. The average current is calculated with an advertising interval of \SI{100}{\milli\second} implemented in the on-board firmware. The implemented chip antenna allows a smaller device with a range of several meters at \SI{0}{\deci\bel} transmit power.

    \subsection{Receiver nodes}
\begin{figure}[htbp!]
            \centerline{\includegraphics[width=0.8\columnwidth]{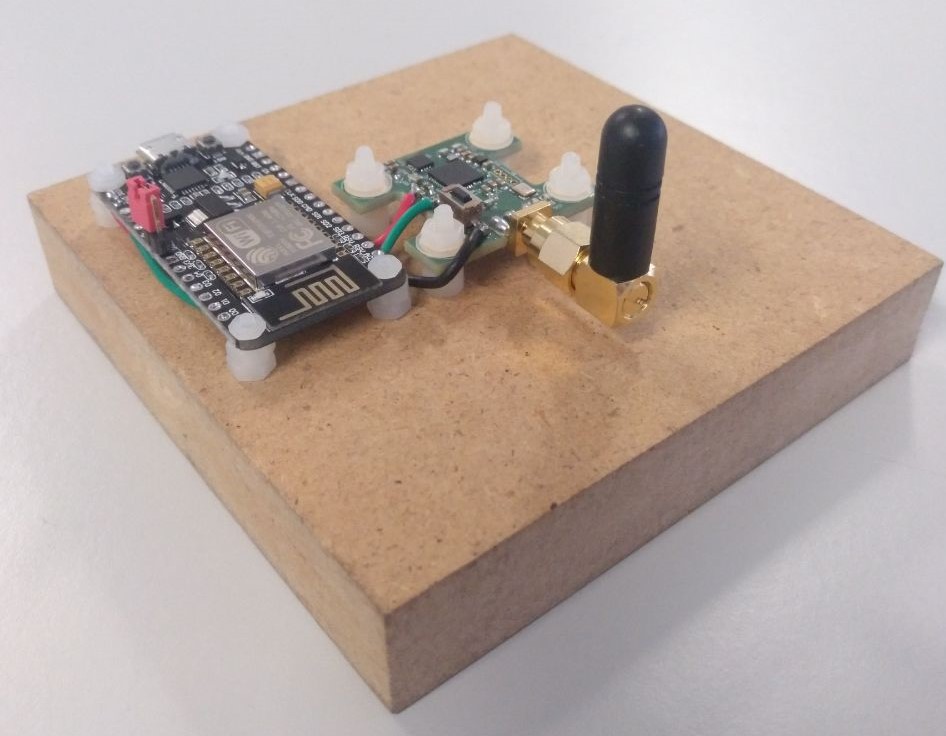}}
            \caption{Prototype of the receiver including the antenna and \acs{WLAN} module.}
            \label{receiver}
        \end{figure}

        In our system, messages from the fixed beacon nodes are received by a mobile receiver. The receiver (Fig.~\ref{receiver}) was designed using the same architecture of the beacon node and therefore has the same components as the beacon. The only differences are the elimination of the battery compartment and its replacement with a \acs{USB}-to-\acs{UART} bridge that also serves as an interface to an external device. To obtain results that are less dependent on the orientation of the antenna, the chip antenna was replaced with an external antenna and the \(\pi\)-network was adjusted accordingly. To obtain the best possible results from the \ac{RSSI} measurements, a holding device was made for the receiver. This consists of a small piece of wood on which the receiver is fixed with screws made of plastic and not steel, as these could potentially cause inaccuracies. With the help of a \ac{WLAN} module that sends information to an access point, the receiver can be used on its own equipped with a power bank.

%% file: content/04_algorithm.tex
\section{\Acl{wkNN} Indoor Localization Algorithm}
\begin{figure}[htbp!]
        \centerline{\includegraphics[width=0.8\columnwidth]{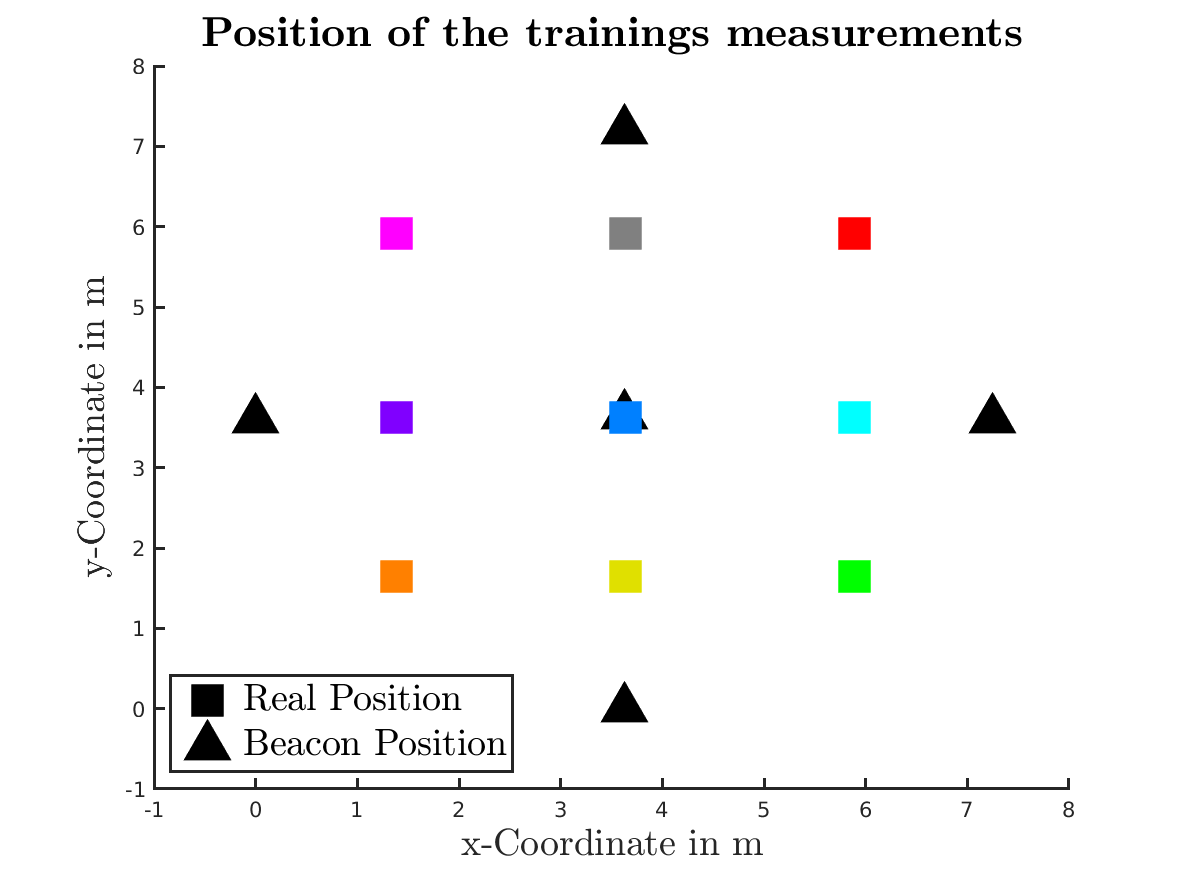}}
        \vspace{-0.5cm}
        \caption{Experimental setup used to acquire training data. The setup consists of 5 beacons and 9 positions where the mobile receiver can be placed to get data from known position.}
        \label{experimental_setup}
    \end{figure}

    The main contribution of this paper is the design and implementation of the \acf{wkNN} algorithm with \ac{RSSI} information using the nRF52832 for training and evaluation.  The \ac{wkNN} algorithm is one of the most widely used algorithms to improve the performance of indoor localization based on \ac{RSSI}~\cite{33}. Moreover, the algorithm runs on the ARM Cortex-M4F microcontroller in the nRF52832 because it does not require large memory and computational resources. This algorithm requires that the position of the beacons and the positions of the training points are known. Fig.~\ref{experimental_setup} shows the setup used to acquire the data set for training.  Through a previous training (fingerprint), the \ac{RSSI} values at each position are determined (vector with one \ac{RSSI} value per beacon). Due to the nature of \ac{wkNN}, only positions within the trained positions can be estimated.
    \par
    \Ac{wkNN} is a complement to \ac{kNN}. \Ac{kNN} attempts to match the measured \ac{RSSI} values with the previously trained positions by finding the distance between all measured \ac{RSSI} values of the training position using the Chebyshev or Euclidean norm.
    \begin{figure}[htbp!]
        \centerline{\includegraphics[width=0.8\columnwidth]{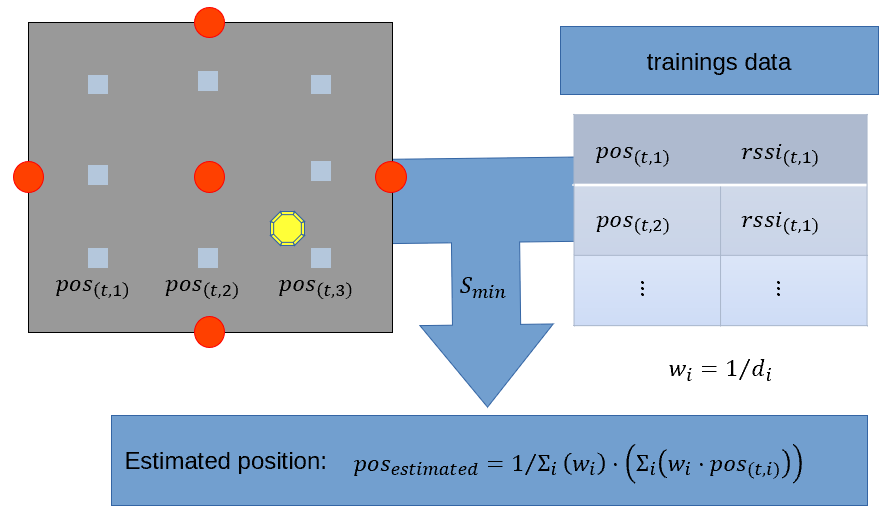}}
        \vspace{-0.4cm}
        \caption{The overview of the \acs{wkNN} with the used application scenario (left).}
        \label{algorithm}
        \vspace{-0.5cm}
    \end{figure}
    \par
    Then \(k\) smallest differences are taken to form an arithmetic mean, which then corresponds to the estimated position.  In \ac{wkNN}, the \(k\) closest training data with their positions are considered. The difference with \ac{kNN} is that in \ac{wkNN}, the arithmetic mean is formed by weighting the position of the \(k\) nearest neighbors by the reciprocal of the estimated distance to the receiver. This assigns a very high value to the training position with a difference close to zero, while the others split the remainder, promoting a convergence to the nearest positions. According to this value, the averaging is now done in favor of the correct position. It is important to note that only positions within the area spanned by the positions of the training data can be correctly detected by averaging.

%% file: content/05_evaluation.tex
\section{Experimental Evaluation}
\begin{figure}[htbp!]
        \centerline{\includegraphics[width=0.8\columnwidth]{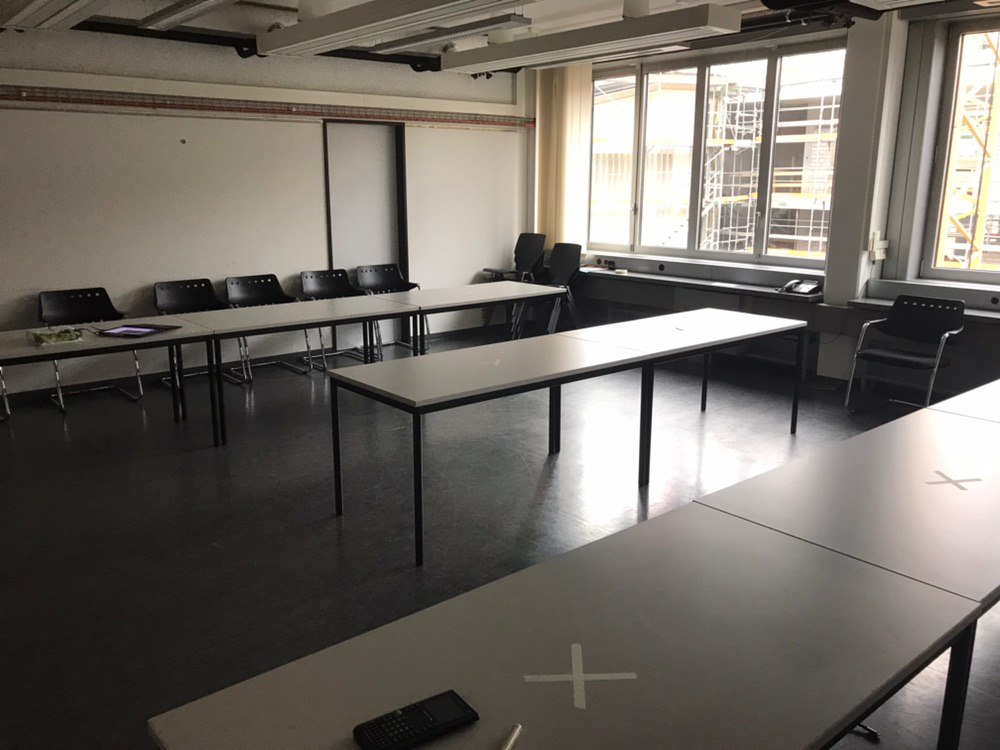}}
        \vspace{-0.25cm}
        \caption{The classroom used for experimental evaluation had an area of \(\SI{7.2}{\meter}\times\SI{7.2}{\meter}\), with the beacons at a height of 1.8m. Obstacles in the room were the chairs and tables, which were approximately 1m high.}
        \label{experimental_room}
    \end{figure}

    To evaluate the designed prototypes and algorithm, we distributed four beacons with the setup shown in Fig.~\ref{experimental_setup}. Fig.~\ref{experimental_room} shows the classroom used for the experimental evaluation with an area of \(\SI{7.2}{\meter}\times\SI{7.2}{\meter}\). The beacons were placed at a height of \SI{1.8}{\meter}. Obstacles in the room were the chairs and desks, which were about 1m high, as shown in the figure.
    \par
\begin{figure}[htbp!]
        \centerline{\includegraphics[width=0.8\columnwidth]{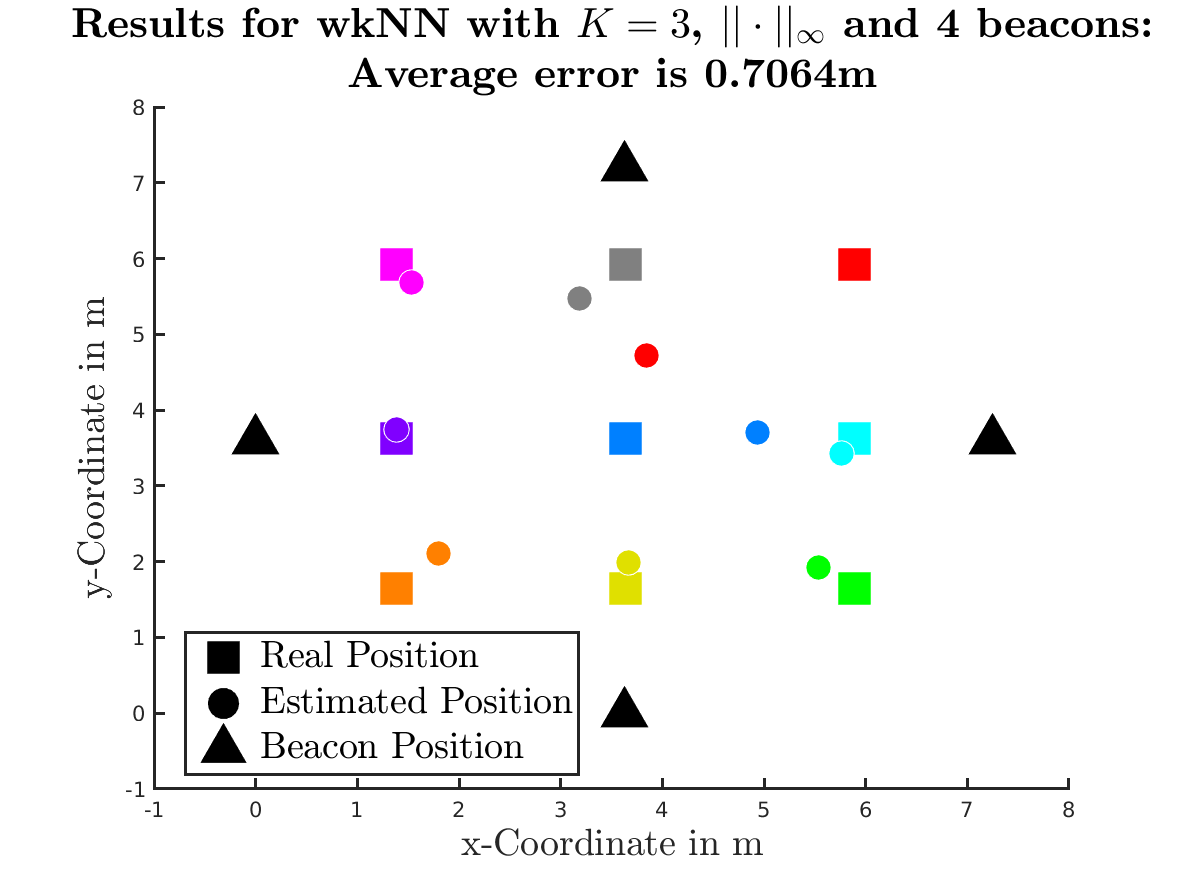}}
        \vspace{-0.5cm}
        \caption{Training positions compared to real-time measurements calculated with the implemented \acs{wkNN} with \(k=3\), Chebyshev norm.}
        \label{k3_4b_max}
    \end{figure}
\begin{figure}[htbp!]
        \centerline{\includegraphics[width=0.8\columnwidth]{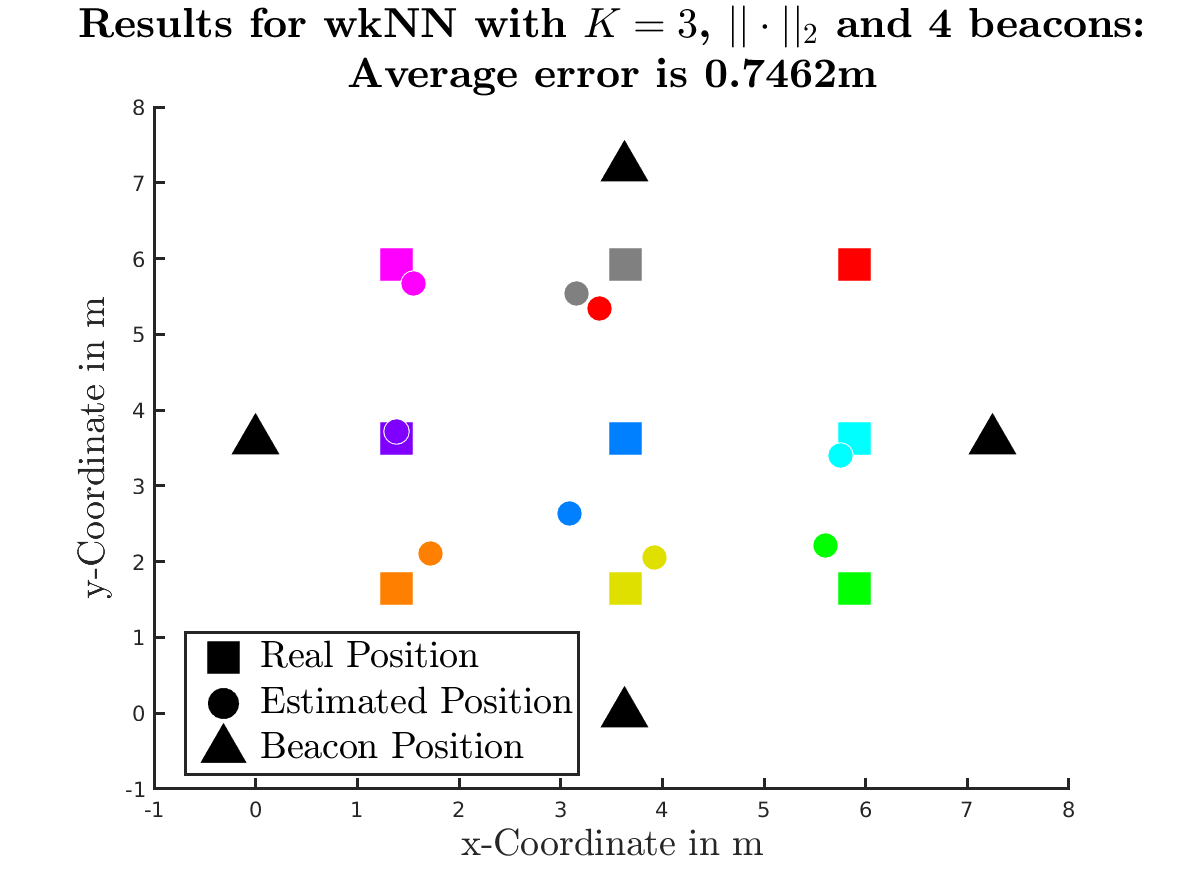}}
        \vspace{-0.5cm}
        \caption{Training positions compared to real-time measurements calculated with the implemented \acs{wkNN} with \(k=3\), Euclidian norm.}
        \label{k3_4b_euc}
    \end{figure}
\begin{figure}[htbp!]
        \centerline{\includegraphics[width=0.8\columnwidth]{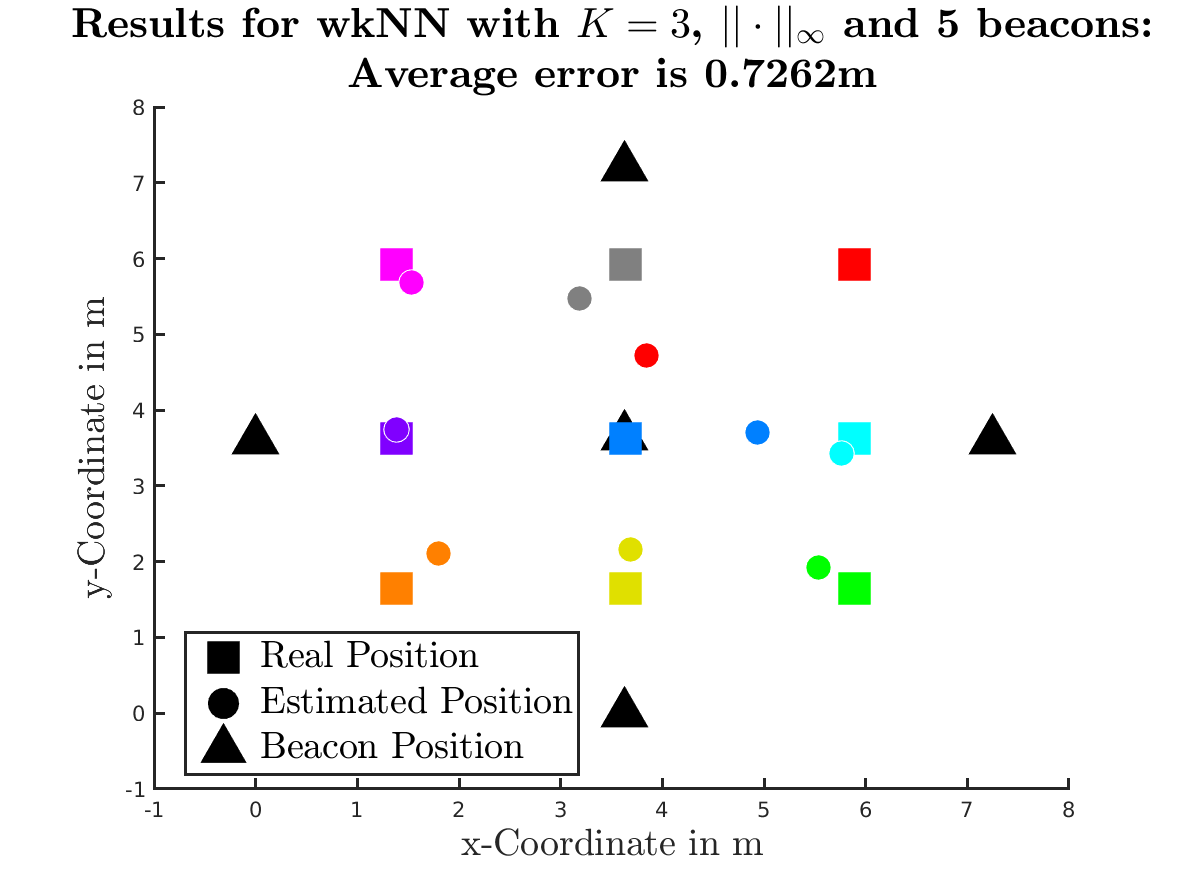}}
        \vspace{-0.5cm}
        \caption{Training positions compared to real-time measurements calculated with the implemented \acs{wkNN} with \(k=3\), Chebyshev norm and 5 beacons.}
        \label{k3_5b_max}
    \end{figure}

    It is important to note that the implemented algorithm is evaluated as a regression problem and not as a classification problem of the 9 known positions. This is done to obtain a more realistic and useful system that can be used in an indoor localization system for medical applications. However, to obtain an evaluation of the accuracy of the estimation, the reference positions were created identical to the trained positions. The finally used value for \(k\) resulted from previous experiments with different parameter choices. Fig.~\ref{k3_4b_max} shows the average error obtained over all 9 positions using the Chebyshev norm, \(k=3\) and only 4 of the five beacons. The average error is only \SI{0.704}{\meter} over 1000 samples. In the experimental setup, the accuracy was found to be \SI{2.5}{\meter} in the worst case and \SI{0.27}{\meter} in the best case. Fig.~\ref{k3_4b_euc} shows instead that the Euclidean norm performs slightly worse in terms of average error, at \SI{0.746}{\meter}. On the other hand, the reported maximum error was lower than \SI{2.37}{\meter}. To evaluate the influence of the fifth beacon we re-evaluated the best configuration with Chebyshev norm, \(k=3\) and 5 beacons. As Fig.~\ref{k3_5b_max} shows, the performance deteriorates with increasing number of beacons, which is due to the asymmetric \ac{RSSI} of the central beacon.
    \vspace{-0.2cm}
    \vspace{-0.5cm}
    \vspace{-0.5cm}
    \vspace{-0.2cm}

%% file: content/06_conclusion.tex
\section{Conclusion}
    This paper presents the design and implementation of an indoor localization hardware-software system that enables localization and tracking in medical applications. The algorithms use a low-complexity algorithm based on \ac{RSSI} and \acl{wkNN} that can run on an ARM Cortex-M4F microcontroller. Experimental results have shown high accuracy of the solution with an average error of only \SI{0.7}{\meter} in a \(\SI{7.2}{\meter}\times\SI{7.2}{\meter}\) scene. The average power consumption of the mobile tag is only \SI{50}{\micro\ampere} at \SI{3}{\volt}, resulting in a long lifetime when powered by a coin battery. The advantages of the presented system are the availability of \acs{BLE} on many conventional devices, the high energy efficiency and the reliable use for rough localization at least room by room. Future work will improve the experimental evaluation and find the optimal parameters of the algorithm.